\documentclass{icrc2009}

\usepackage{graphicx}   % for including figures
\usepackage{caption}    % for captions
\usepackage[font=footnotesize]{subfig} % subfig.sty for a double column floating figure using two subfigures
\usepackage{fixltx2e}
\usepackage{url}

\newcommand{\shorttitle}[1]%
{\markboth{Proceedings of the 31\MakeLowercase{$^{st}$} ICRC, {\L}\'{o}d\'{z} 2009}{#1} }
\newcommand{\etal}{\MakeLowercase{\textit{et al. }}} % "et al."

\newcommand{\ga}{\gamma}
\newcommand{\beq}{\begin{equation}}
\newcommand{\eeq}{\end{equation}}
\newcommand{\ove}{\overline}
\newcommand{\eps}{\varepsilon}

\begin{document}
\title{Search for neutrino bursts from core collapse supernovae
at the Baksan Underground Scintillation Telescope.}

\author{\IEEEauthorblockN{R.V. Novoseltseva, M.M. Boliev, I.M. Dzaparova,
              M.M. Kochkarov, S.P. Mikheyev,\\
              Yu.F. Novoseltsev, V.B. Petkov, P.S. Striganov, G.V. Volchenko,
              V.I. Volchenko, A.F.
              Yanin\IEEEauthorrefmark{1}}
                            \\
\IEEEauthorblockA{\IEEEauthorrefmark{1}Institute for Nuclear
        Researsh of the Russian Academy of Sciences,\\
        60-th October Anniversary prospect 7a, Moscow 117312, Russia }}

% please write the preseter's name and short title (3-4 words maximum)
%    which will appear at the header of the even pages.
\shorttitle{R.V. Novoseltseva \etal Search for neutrino bursts ...}
\maketitle

\begin{abstract}
Current status and results of the experiment on recording neutrino
bursts are presented. The observation time (since 1980) is 24.7
years. The upper bound of collapse frequency in our Galaxy is 0.093
$y^{-1}$ (90\% CL). \vspace{1pc}
  \end{abstract}


\section{Introduction}
One of the current task of the Baksan Underground Scintillation
Telescope (BUST) is the search for neutrino bursts from
gravitational collapse of stars.

BUST is located in the Northern Caucasus in the underground
laboratory at the effective depth of $8.5\times 10^4\ g\cdot
cm^{-2}$ (850 m of w.e.) \cite {Alex}. The facility has dimensions
$17\times 17\times 11$ m$^3$ and consists of four horizontal
scintillation planes and four vertical ones (Fig.~1). Five planes of
them are external planes and three lower horizontal planes are
internal ones. The upper horizontal plane consists of 576 ($24\times
24$) liquid scintillator detectors of the standard type, three lower
planes have 400 ($20\times 20$) detectors each. The vertical planes
have $15\times 24$ and $15\times 22$ detectors. The detector sizes
are $0.7\times 0.7\times 0.3\ m^3$. The distance between neighboring
horizontal scintillation layers is 3.6 m. The angular resolution of
the facility is 2$^o$, time resolution is 5 ns.

 The information from each detector is transmitted
over three channels:  an anode channel (which serves for trigger
formation and amplitude measurements up to 2.5 GeV), a pulse channel
with operation threshold 12 MeV (since 1991 this threshold = 8 MeV;
the most probable energy deposition of a muon in a detector is 50
MeV $\equiv $ 1~relativistic particle) and a logarithmic channel
with a threshold $s_o = 0.5$ GeV.  The signal from the fifth dynode
of PM tube FEU-49 goes to a logarithmic
 channel (LC) where it is converted  into a pulse whose
length t is proportional to the logarithm of the amplitude of the
signal \cite {Bak}.
 \begin{figure}[!t]
  \centering
  \includegraphics[width=3in]{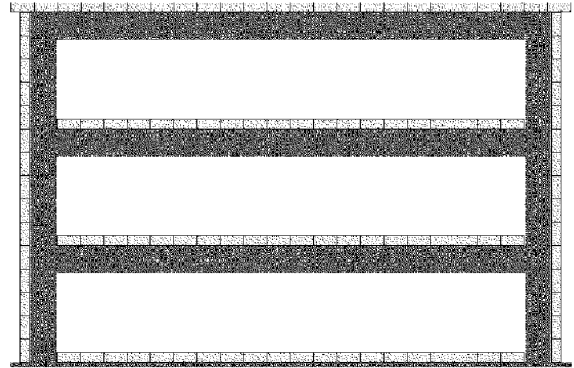}
  \caption{General view of the Baksan underground
           scintillation telescope.}
  \label{fig:fig1}
 \end{figure}

There is a voluminous set of monitoring programs which allows us to
examine the operation of each detector of the facility in any time
interval and discard not properly working detectors.

\section{The method of neutrino burst detection}
BUST consists of 3156 standard autonomous detectors. Each detector
is filled with an organic scintillator $C_nH_{2n+2}$, ($n\simeq 9$)
and is viewed by one photomultiplier with a photocathode diameter of
15 cm. The total scintillator mass is 330 t, and the mass enclosed
in three lower horizontal layers (1200 standard detectors) is 130
tons. The neutrino signal from a supernova explosion is recorded
with a help of a reaction
\begin{equation}
\ove \nu_e + p \to n + e^+ \label {1}
\end{equation}
If the mean antineutrino energy is $E_{\nu _e} = 12 - 15$ MeV \cite
{Imsh1,Hill,Imsh2} the pass of $e^+$ (produced in reaction (\ref
{1})) will be included, as a rule, in the volume of one detector. In
such case the signal from a supernova explosion will appear as a
series of events from singly triggered detectors (one and only one
detector from 3156) during the neutrino burst. The search for a
neutrino burst consists in recording of single events bunch within
time interval of $\tau $ = 20 s.

If one assumes the distance from the star is 10 kpc and the total
energy irradiated in neutrinos is
\begin{equation}
\eps_{tot} = 3\times 10^{53} \ erg \label {2}
\end{equation}
the expected number of single events from reaction (\ref {1}) (we
assume the total energy of the $\ove \nu_e$ flux is equal to
$1/6\times \eps_{tot}$) will be
\begin{equation}
N^H_{ev} \simeq 38\times \eta_1, \label {3}
\end{equation}
where $\eta_1$ denotes the detection efficiency of $e^+$ in reaction
(\ref {1}) and the symbol "H" indicates that the hydrogen is the
target. $\eta_1\approx $ 0.7 if the electron energy $E_e =$ 10 MeV
and $\eta_1 = $ 0.9 if $E_e =$ 20 MeV.

Background events are radioactivity, ghost signals from detectors
and cosmic ray muons if only one detector from 3156 hit. The total
count rate from background events is n = 0.02 $s^{-1}$ in internal
planes (three lower horizontal layers) and 1 $s^{-1}$ in external
ones. Therefore three lower horizontal layers are used as a target
(the estimation (\ref {3}) has been calculated for three internal
planes).

Background events can imitate the expected signal (k single events
within sliding time interval $\tau$) with a count rate
\begin{equation}
p(k) = n\times exp(-n\tau)\frac{(n\tau)^{k-1}}{(k-1)!} \label {100}
\end{equation}
The treatment of experimental data (background events over a period
2001 - 2008 y; T = 236126 RUNs, RUN = 900 s) is shown by squares in
Fig.2 in comparison with the expected distribution according to the
expression (\ref {100}).

If the scenario of 2-stage collapse \cite {Imsh3} is realized in
Nature and the mean neutrino energy (during the first stage) is
$\ove E_{\nu _e} = 30 - 40$ MeV, the following reactions begin to
work:

\begin{equation}
\begin{array}{l}
\nu_i + ^{12}C \to ^{12}C^* + \nu_i, \hspace{5mm} E_{th}=15.1\ MeV,
\\[2mm] \ i = e,\mu , \tau,
\\[3mm]
^{12}C^* \to ^{12}C + \ga, \hspace{5mm} E_{\ga} = 15.1\ MeV
\end{array}
\label{4}
\end{equation}
%$$^{12}C^* \to ^{12}C + \ga, \hspace{5mm} E_{\ga} = 15.1\ MeV$$
and
\begin{equation}
\begin{array}{l}
\nu_e + ^{12}C \to ^{12}N + e^-, \hspace{5mm} E_{th}=17.34\ MeV,
\\[3mm]
^{12}N \to ^{12}C + e^+ + \nu_e, \hspace{5mm}\tau = 15.9\ ms,
\end{array} \label {5}
\end{equation}
\noindent $\tau$ is a lifetime of the nucleus $^{12}N$.

Reaction (\ref {4}) allows to measure the total neutrino flux with
the energy $E_{\nu} > 15.1$ MeV.

If the mean energy $\ove E_{\nu}=$ 30 MeV the expected number of
events for reactions (\ref {4}) and (\ref {5}) can be estimated
(under conditions (\ref {2})) by formulae
\begin{equation}
N^C_{ev2} = 16\times \eta_2(E_{\ga} = 15\ MeV), \label {6}
\end{equation}
\begin{equation}
N^C_{ev3} = 30\times \eta_3(E_{\nu} = 30\ MeV), \label {7}
\end{equation}
The radiation length for our scintillator is 47 g/cm$^2$ therefore
$\eta_2 \approx 0.2$.  In reaction (\ref {5}) BUST can detect both
$e^-$ with energy $E_{\nu} - 17$  MeV and $e^+$ if the energy
deposition from these particles is greater 8 MeV. In the latter
case, the reaction (\ref {5}) will have the distinctive signature:
two signals separated with $5 - 45$ ms time interval (dead time of
the BUST is $\simeq 1$ ms).

 In reaction (\ref {5}) the sum of energies $E_{e^+} + E_{\nu}$ is 17.3 MeV therefore
$\eta_3 \approx 0.5 - 0.7$.

The low part of the overlap between horizontal scintillation planes
is the 8 mm iron layer. This can be used as the target in the
reaction
\begin{equation}
\nu_e + ^{56}Fe \to ^{56}Co^* + e^-, \hspace{2mm} E_{th}=4.056\ MeV,
\label {Fe}
\end{equation}
(cobalt emerges in excited state).

Under conditions (\ref {2}) the expected number of events from
reaction (\ref {Fe}) (neutrinos arrive from above) is
\begin{equation}
N^{Fe}_{ev} = 6.3\times \eta_{Fe}(26\ MeV), \label {9}
\end{equation}
$\eta_{Fe}(26\ MeV) \approx 0.4$ is the detection efficiency of
$e^-$ with the energy 26 MeV produced into the 8 mm iron layer.

It should be noticed, if $\ove E_{\nu _e} = 30 - 40$ MeV a
noticeable percentage of neutrino reactions (\ref {Fe}) will cause
triggering two detectors.
 \begin{figure}[t]
  \centering
 \hspace*{-8mm}
  \includegraphics[width=4in]{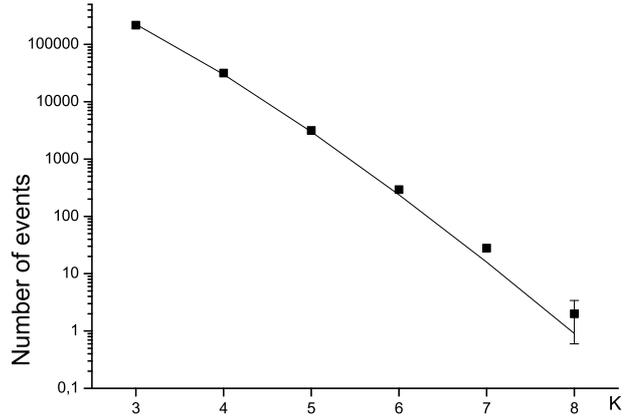}
\caption{The number of bunches with $k$ single events within time
interval of $\tau$ = 20 s. Squares are experimental data, the curve
is the expected number according to the expression (\ref {100}).}
\label{fig:fig2}
\end{figure}

Since 2001 all events are collected by the facility. Earlier only
the events selected by physics programs (by means of corresponding
electronic systems) were recorded. All events recording allows us to
observe any events before and after single events.

\section{Results}

So, if the scenario of 2-stage collapse \cite {Imsh3} is realized in
Nature the signal from collapse (the number of neutrino induced
events) is increased $\approx 50\%$.

%As can be seen above ...

The facility has been operating under the program of search for
collapse neutrinos since the mid-1980 \cite {Alex2}. The observation
time is $T=$ 24.7 years. Let $f_{col}$ be the mean frequency of
collapses. The probability of collapse absence during the time
interval $T$ is (according to the Poisson law) $\exp(-f_{col}T)$. An
upper bound on the mean frequency of gravitational collapses in the
Galaxy at 90\% CL can be obtained with the help of the expression
\begin{equation}
\exp(-f_{col}T) = 0.1
\end{equation}
Thus
\begin{equation}
f_{col} < 0.093 y^{-1},\ \ \ 90\% \ CL
\end{equation}

{\bf Acknowledgements.} The work is supported by Russian Fund for
Basic Research (grants no. 07-02-00162 and 09-02-00434), Russian
Academy of Sciences Basic Research Program "Neutrino Physics and
Astrophysics" and State Program of Supporting Leading Scientific
Schools (grant no. NSh-321.2008.2).

\end{document}